\documentclass[final]{raa07}            

\usepackage{graphicx,times}             
\usepackage{natbib}
\usepackage{amssymb,amsmath}
\usepackage{mathrsfs}
\bibpunct{(}{)}{;}{a}{}{,}
\renewcommand{\thepage}{156--\arabic{page}}
\begin{document}

\title{SETI {strategy}
{with} FAST fractality}

\volnopage{{\bf 2021} Vol.~{\bf 21} No.~{\bf 7},~178(7pp)~
   {\small  doi: 10.1088/1674-4527/21/7/178}}
   \setcounter{page}{1}         

   \author{Yi-Xuan Chen
      \inst{1}
      \and Wen-Fei Liu
      \inst{2}
   \and Zhi-Song Zhang
      \inst{3}
   \and Tong-Jie Zhang
      \inst{1,2}
   }

  \institute{Department of Astronomy, Beijing Normal University, Beijing 100875, China;
         {\it tjzhang@bnu.edu.cn}\\
         \and College of Physics and Electronic Engineering, Qilu Normal University, Jinan 250200, China\\
         \and
          National Astronomical Observatories, Chinese Academy of Sciences, Beijing 100101, China\\
 \vs\no
 {\small Received~~2021 February 17; Accepted~~2021~~April 12}
 }
\abstract{We {applied} the Koch snowflake fractal antenna in
{planning calibration} of{ the} Five-hundred-meter Aperture
Spherical radio Telescope (FAST), {hypothesizing} second-order
fractal primary reflectors {can} optimize the orientated sensitivity
of the telescope. Meanwhile, on the grounds of NASA Science Working
Group Report in 1984, we reexamine the strategy of Search for
Extraterrestrial Intelligence (SETI). {A} mathematical analysis of
the radar equation will be {performed} in the first section,
aim{ing} to make it convenient to design a receiver system that can
{detect} activities of an extraterrestrial civilization, according
to the observable region of the narrowband. Taking advantage of the
inherent potential of FAST, we simulate the theoretical detection
of{ a} Kardashev Type I civilization by{ a} snowflake-selected
reflecting area~(\citeauthor{paper:04}).  }

   \authorrunning{{\it Y.-X. Chen et al.}: SETI Tactics on FAST Fractality }            
   \titlerunning{{\it Y.-X. Chen et al.}:  SETI Tactics on FAST Fractality}  
   \maketitle

%
\section{Overview}          
\label{sect:intro}

{At} the outset, Dunning et al. put forward {a} blueprint for an L-band
orientation cryogenic 19 multi-beam feed system~\citep{paper:05} in
August 2017. The Stewart platform cabin~\citep{paper:06} {represented} an
improvement {over the} CSIRO Parkes Observatory at the Australian
National Telescope Facility (ATNF). In recent
years, the scientific work {at} National Astronomical
Observatories{, Chinese Academy of Sciences} (NAOC) has gradually
expanded to {possible} recognition of extraterrestrial
signals~\citep{paper:08}. In this section, we briefly review the
status quo {in} application of{ a} {Search for Extraterrestrial
Intelligence (}SETI{)} theoretical framework to antenna design to
provide a direct description of the fractal {refinement} of{ the}
{Five-hundred-meter Aperture Spherical radio Telescope (}FAST{)}.

We{ start by} introduc{ing} the holomorphic Helmholtz
equation for {a} radiat{ing}
antenna \begin{equation} \Delta u+k^2u=0\,.
 \label{1}
\end{equation}

Inspired by the target detection method mentioned in{ the}
Exotic Target Catalog~\citep{paper:10}, we let function $s$
represent the probe ~\citep{paper:09} for the stereoscopic
survey (Appendix A) \begin{equation}
u(d,\theta,\varphi)=\frac{{\rm
e}^{-jR}}{R}\sum_{n}^{\infty}\frac{s_n(\theta,\varphi)}{R^n}\,,
 \label{2}
\end{equation} where the denotations $\theta, \varphi$ are angular
variables of the radiation. Based on extrapolating elliptical
radiation characteristics of antenna detection in Appendix A, we
introduce the polarized voltage {\textit{V}} of
the antenna receiver as ideal piezoelectricity
\begin{equation}
{\textit{V}}(t)=\sum_{n}^{\infty}|C_{n}|{\rm
e}^{j2\pi v_{n}t}, \forall~C\in\mathfrak{R}^+\,.
 \label{3}
\end{equation} In order to scale the polyphaser in
Equation~\eqref{3}, we estimate the  power spectrum {in}
the Nyquist form. Suppose the density of the noise voltage across a
resistor with complex impedance $Z$ is in thermal
equilibrium~\citep{paper:11} \begin{equation} {\rm
d\overline{{\textit{V}}^2}}=4dk_BT_{\rm
sys}v\cdot {\rm re}\{Z\}\left[0.5+\frac{1}{{\rm e}^{(hv/k_BT_{\rm
sys})}-1}\right]{\rm d}v{.}
 \label{4}
\end{equation} {W}e claim the simplified excess noise
ratio (ENR) from the time average for the band-pass filter
\begin{equation}
\overline{\dot{{\textit{V}}}}\approx
4dk_{B}T_{\rm sys}\Delta v=4d\times{\rm ENR}\,,
 \label{5}
\end{equation} where the equivalent noise bandwidth (ENBW) has been
established (Appendix B). To extend the expression
of the noise factor $f$, we reference one of the applications for
ideal {intermediate frequency (}IF{)} sampling receivers
({e.g.} {t}he dual high-speed differential
amplifier LTC6420-20 targeted at processing signals from DC to
300{\,}MHz), described{ by} the term \begin{equation}
f=f_{1}+\sum_{k=2}^{m}f^k\left(\prod_{i=1}^{m-1}G_{i}\right)^{-1}\,,
\label{6} \end{equation} where gain $G$ and the series{ is
defined} as $\{f^k\triangleq f_{k}-1|k=2,3,...,m\}$, and the
relationship with the signal-to-noise ratio (SNR) is
dimensionless \begin{equation} f=\rm SNR_{in}/SNR_{out}\,.
\label{7} \end{equation} On the other hand, we define $P_{\rm on}$
and $P_{\rm off}$ {to be} the power level of the
receivers measured {at} the moment of{ being}
switch{ed} on and off{ respectively}, following the ratio
\begin{equation} \eta_{\rm sw}=P_{\rm on}/P_{\rm off} \label{8}
\end{equation} to do logarithmic processing of the noise figure
(NF) \begin{equation} {\rm
NF}=10\log_{10}f=10\log_{10}\left(\frac{10^{\rm
ENR/10}}{10^{\eta_{\rm s}/10}-1}\right) \label{9} \end{equation}
concluding \begin{equation} {\rm NF}={\rm ENR}-10\log_{10}(\eta_{\rm
s}-1)\,. \label{10} \end{equation} For the photoelectric conversion
of the feed cabin, we {express} the {NF} {by}
considerin{g} temperature
\begin{equation}
\rm\frac{SNR_{in}}{SNR_{out}}=1+\frac{\emph{T}_{R}}{\emph{T}_{o}}
\label{11} \end{equation} to substitute into the gain control that
under the steady state intermediate frequency amplifier
(IFA)~\citep{paper:12} {yields}
\begin{equation} 1+{\rm Y}=\frac{\rm ENR}{\eta_{\rm
s}-1}\,{.} \label{12} \end{equation} {Here}, we set factor Y as the Johnson-Nyquist thermal
noise representation of FAST from Equation~\eqref{11}
\begin{equation} {\rm Y}=T_{\rm R}/T_{\rm o}\,, \label{13}
\end{equation} yielding the noise temperature of the receiver
\begin{equation} T_{\rm R}=T_{\rm o}\left(\frac{\rm
ENR}{\rm\eta_s-1}-1\right) \label{14} \end{equation} at $T_{\rm
o}=290{\,}{\rm K}$. Meanwhile, we consider the system
loss{;} for all isotropic antennae, the system gain that
{occurs with} the initial temporal
distancing must be estimated \begin{equation} G_{\rm
sys}=P_{\rm in}-P_{\rm re}+L_{\rm sys}\,. \label{15} \end{equation}
After {rearranging} both sides of the
equation, we have \begin{equation} \begin{aligned}
P_{\rm in}-P_{\rm re}&=|G_{\rm sys}-L_{\rm sys}|\\
&=10\log_{10}\left[\eta\left(\frac{\pi d}{\lambda}\right)^2\right]\!-\!20\log_{10}\Delta v\!\\
&\quad -\!20\log_{10}d\!-\!35.77, \end{aligned} \label{16}
\end{equation} where $P_{\rm re}$ denotes the interference signal
level. We recall the noise power \begin{equation} {\rm
NF}=kT\Delta v_{\rm noise}\,, \label{17} \end{equation} which is
standardized {at} 1{\,}Hz, the back end of FAST is
at room temperature (293{\,}K) and the contribution to
white noise is $-$174 dBm. In
{\cite{paper:13}}, for
system gain{ with} respect to receivers, the expression
can be describe{d} as below \begin{equation}
\begin{aligned}
G_{\rm sys}=&P_{\rm on}-{\rm NF}-10\log_{10}({\rm ENR}+1)\\
&-10\log_{10}\Delta v+174\,, \label{18} \end{aligned} \end{equation}
where the antenna gain within{ a} zenith angle of $\pi/6$
is 54.4$\sim$74.1{ }(dB) on average, and the sensitivity is
ideally about 1600{\,}$\rm(m^2{\,}K{^{-1}})$
{if} the efficiency of the telescope is 57\%. On the
other hand, Equation~\eqref{17} can be
{decomposed} as the level of noise on average
\begin{equation} P=k_{B}T\Delta v \label{19} \end{equation} while we
{compute} logarithm on both sides \begin{equation}
\log_{10}P\cong \log_{10}(k_{B}T)+\log_{10}\Delta v\,, \label{20}
\end{equation} which{ is} directly proportional to the bandwidth
$P\propto\Delta v$. For the purpose of
{measuring} the sensitivity F, the system
equivalent flux density (SEFD) must be introduced as a bridge
\begin{equation} {\rm SEFD}=\frac{2k_{B}T_{\rm sys}}{\eta A}\,,
\label{21} \end{equation} where the physical collecting area of the
telescope{ is} $A$ and dimensionless efficiency factor
$\eta\in[0,1]${ is} normally between 0.5 $\sim$ 0.7
without distortion. From the perspective of a radio telescope, we define the
effective collecting area \begin{equation}
\eta\frac{G\lambda^2}{4\pi}=\eta A\equiv A_{\rm eff}\,. \label{22}
\end{equation} In addition, according to {experience studying} military radars and TV
station antennas, M.~Zaldarriaga \& A.~Loeb
claim that the point source sensitivity (PSS) of an
interferometer composed of antennae may {collect radiation that could be detected by} SETI with
redshifted 21{\,}cm observation
\begin{equation} {\rm F_{PSS}}={\rm costant}\times\frac{\rm
SEFD}{\sqrt{\Delta vt_{\rm ob}N_{\rm b}}}\,. \label{23}
\end{equation} However, to estimate the constant, we assume that the
receiver is optimized to ensure the SNR{ is reasonable}. In the first
step, the frequency domain criterion of the filter is defined by the
complex conjugate spectrum. In other words, the contribution of the
spectral function to SEFD can be expanded to \begin{equation}
\begin{aligned}
{\rm SNR}(\tau)&=\frac{\left[\int_{-\infty}^{+\infty}|\mathcal{S}(v)|^2{\rm d}v\right]^2}{2\pi\varrho\int_{-\infty}^{+\infty}|\mathcal{S}(v)|^2{\rm d}v}\\
&=\frac{1}{2\pi\varrho}\int_{-\infty}^{+\infty}|\mathcal{S}(v)|^2{\rm d}v\overset{\text{def}}{=}{\rm SN{R}},\\
&\quad\quad{\rm when}~~\tau=t, \end{aligned} \label{24}
\end{equation} where $\varrho$ represents the spectr{al}
power density from the background of stationary noise $\varrho$.
Return{ing} to Equation~\eqref{23},
we adjust the PSS equation \begin{equation} \fbox{${\rm
F}={\rm SNR}\cdot\frac{\rm SEFD}{\sqrt{\Delta vt_{\rm ob}N_{\rm
b}}}$}\,, \label{25} \end{equation} {for }the baseline
$N_{\rm b}$ of sweeps that {are }observe{d} for a
time $t_{\rm ob}$ with a bandwidth $\Delta v$. {W}e completed the proof
{f}o{r} the sensitivity F. Consider{ing} the
pulse repetition frequencies (PRF) with constant observation time,
we {find} the relationship \begin{equation}
{\rm S_{Pmin}}\propto \frac{1}{\sqrt{N_{\rm b}}}\,, \label{26}
\end{equation} where the minimal signal power{ is} $\rm
S_{Pmin}$. Given the physical antenna gain \begin{equation}
\begin{aligned}
G=\left(\frac{4\pi}{\lambda^2}\right)A_{\rm eff}&=\left(\frac{4\pi}{\lambda^2}\right)\eta A=\eta\cdot\frac{4\pi}{\lambda^2}\cdot\frac{G_{\rm rec}\lambda^2}{4\pi}\\
&=\eta\cdot\frac{4\pi}{\lambda^2}\cdot\frac{\pi
d^2}{4}=\eta\left(\frac{\pi d}{\lambda}\right)^2 \end{aligned}
\label{27} \end{equation} {where} the gain of the
transmitter has been defined by power density $\varrho$
\begin{equation} G=\frac{\varrho(\theta,\phi)}{P_{\rm tr}/4\pi
d^2}\,, \label{28} \end{equation} we claim the power for the
receiver \begin{equation} P_{\rm rec}={\rm S_{p,rec}}A_{\rm
eff}=\frac{G_{\rm tr}P_{\rm tr}}{4\pi
d^2}\cdot\eta\left(\frac{G_{\rm rec}\lambda^2}{4\pi}\right)
\label{29} \end{equation} so that we {restate} the
Friis transmission for the gradient of its power radiation{
as} \begin{equation} \begin{aligned}
P_{\rm rec}&=\cos^2(vt)\cdot\eta\cdot\frac{G_{\rm rec}G_{\rm tr}P_{\rm tr}c^2}{(4\pi dv)^2}\\
&\equiv{\rm NPF}\cdot\frac{\eta G_{\rm rec}G_{\rm tr}P_{\rm
tr}c^2}{(4\pi dv)^2}, \label{30} \end{aligned} \end{equation} and
the effective radiated power (ERP) that {is
}relate{d} to the signal power within the antenna
\begin{equation} {\rm S_{P}}=\frac{GP}{4\pi
d^2}\cdot\frac{\sigma}{4\pi
d^2}\cdot\frac{G\lambda^2}{4\pi}{.} \label{31}
\end{equation} {C}alculating the {ERP} \begin{equation} {\rm S_{P}}={\rm
ERP}\cdot\frac{\sigma A_{\rm eff}}{16\pi^2 d^4}\,. \label{32}
\end{equation} Return{ing} to Equation~\eqref{31}, we
express the signal power \begin{equation} \begin{aligned}
{\rm S_{P}}&=\left(\frac{1}{4\pi d^2}\cdot\frac{4\pi\eta A}{\lambda^2}\cdot P\right)\left(\frac{\sigma}{4\pi d^2}\right)\left(\frac{1}{4\pi }\cdot\frac{4\pi\eta A}{\lambda^2}\cdot\lambda^2\right)\\
&=\frac{P\sigma\eta^2A^2}{4\pi d^4\lambda^2}\,. \end{aligned}
\label{33} \end{equation} In view of the minimal signal{ having
a} local detection for {recording} the signature,
we obtain the biquadratic distance \begin{equation}
\begin{aligned}
d_{\rm max}^4&\approx\frac{P\sigma GA_{\rm eff}^2}{4\pi\lambda^2{\rm S_{P,min}}}=\frac{P\sigma GA_{\rm eff}^2}{4\pi\lambda^2[P_{\rm int}\cdot f\cdot({\mathop{}_{t\in\mathbb{R}}^{\rm int}\rm SNR_{out}})]}\\
&\resizebox{.9\hsize}{!}{$ =P\sigma A_{\rm
eff}^2/\left[4\pi\lambda^2k_{B}T_{0}\Delta v\cdot\left(\frac{N_{\rm
out}}{Gk_{B}T_{0}\Delta v}\right)\left(\frac{S_{\rm out}}{N_{\rm
out}}\right)_{{\rm min},T}\right]{.} $} \end{aligned}
\label{34} \end{equation} {W}hen the radar temperature is
fixed, the radius has been determined \begin{equation} d_{\rm
max}=\sqrt[4]{\frac{P\sigma A_{\rm eff}^2}{4\pi S_{\rm
P,min}\lambda^2}}\,{.} \label{35} \end{equation} {T}he issue of multiple SNR{ values from} panel
points{ with} R/T components is not the case in this
section. {In addition}, we notice that
the fading depth in the antennae needs to be
{recalled}~\citep{paper:15} \begin{equation}
m=\rm\frac{E^2[RSL]}{Var[(RSL)^2]}\,, \label{36} \end{equation}
which {is related} to received signal level (RSL)
while the average signal power for the antenna in the formula
is $\rm E[RSL^2]$. {\cite{paper:04}} mentioned
that no-slot joint{ exists} to define the
range limitation $d_{\rm max}$ with one degree of freedom
\begin{equation} \begin{aligned}
d_{\rm max}&=\sqrt{\frac{AP}{4\pi mk_{B}T_{0}\Delta v}}\\
&=\sqrt{\left(\frac{\eta\pi l^2}{4}\right)\left(\frac{P}{4\pi mk_{B}T_{0}\Delta v}\right)}\\
&=\sqrt{\frac{l^2\eta P}{16mk_{B}T_{0}\Delta v}}\,{.} \end{aligned}
\label{37} \end{equation} {T}he notation in Eq. \eqref{37}
represent{s} diameter $l\in\mathbb{N}^+$ and aperture efficiency
$\eta$, respectively \begin{equation} d_{\rm
max}=\frac{l}{4}\sqrt{\frac{\eta P}{\pi mk_{B}T_{0}\Delta v}}\,.
\label{38} \end{equation} Taking out the scattering formula for the
receiving power, we ad{o}pt the spheral radiation{ description}
\begin{equation} {\rm EIRP}=\frac{P_{\rm rec}}{G_{\rm rec}}\cdot
\left(\frac{v}{c}\right)^2\cdot(4\pi d)^2. \label{39} \end{equation}
The ratio \begin{equation} {\rm SNR}=\frac{A_{\rm eff}\cdot{\rm
EIRP}}{4\pi d^2k_{\rm B}T_{\rm sys}\Delta v} \label{40}
\end{equation} also satisfies the equation \begin{equation} {\rm
EIRP}=G_{\rm in}+10\log\Delta v\,. \label{41} \end{equation}
Secondar{il}y, we introduce the power of transmitting antenna
$P_{\rm tr}$ so that the distance of the transmitter can be
rewritten as \begin{equation} d_{\rm max}=\sqrt{\frac{G_{\rm
tr}P_{\rm tr}}{4\pi {\rm S_{min}}}}=\sqrt{\frac{\rm EIRP}{4\pi {\rm
S_{min}}}}\,. \label{42} \end{equation} Vice versa,
Equation~\eqref{42} {is }associate{d} with the cut sphere
representation \begin{equation} {\rm EIRP}={\rm S}_{\rm min}(4\pi
d_{\rm max}^2), \label{43} \end{equation} which is equivalent to the
sensitivity \begin{equation} \fbox{${\rm EIRP}={\rm F}\cdot4\pi
d_{\rm max}^2$}\,. \label{44} \end{equation} Considering the actual
performance of non-solid transmitters, the EIRP calculation of FAST
must be bypassed. {F}or the scheme design of{ the} SETI task, we
emphasize the transmission efficiency with the high-level block
diagram of the ROACH board development (e.{g} Xilinx Virtex 6 of{ a}
field-programmable gate array (FPGA) or ADC1X26G layout) from the
FAST backend~\citep{paper:16}, which {can} also detect prebiotic
molecules.

\begin{figure} \centering
\includegraphics[width=0.5\textwidth]{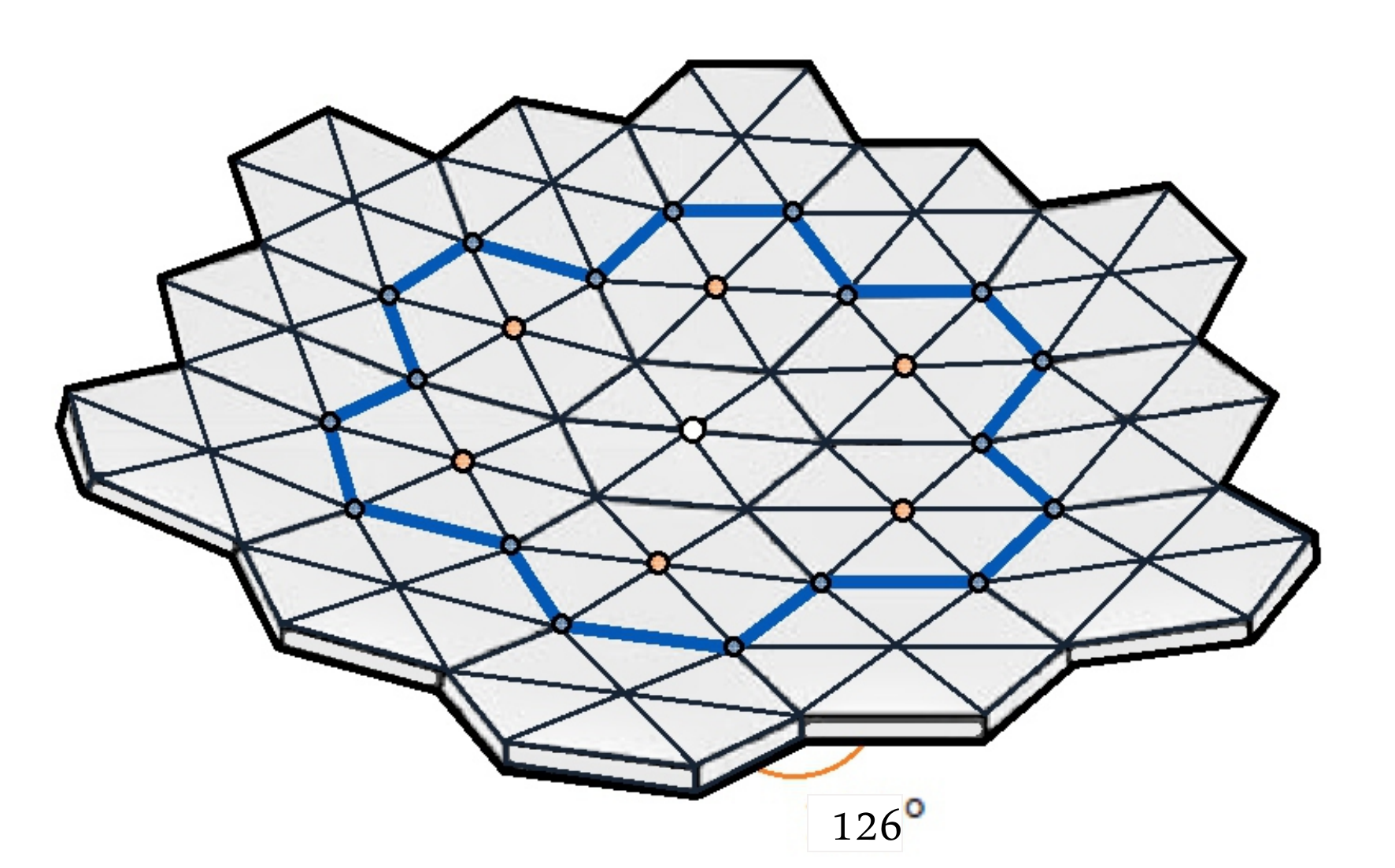}
\caption{\baselineskip 3.8mm A part of the local fractalized main
refl{e}ctor{.} {This}
displays iteration $n=2$. The cente{r} of the
aperture is the 19 mu{l}tibea{m} feed located
{at} the coordinate (498956.8651, 2838440.713,
838.32178).}
\label{1} \end{figure}

\section{Fractalization of the aperture} Consider{ing} that FAST {incorporates} the Gregorian antenna{ design} and
its reflector{ is} composed of perforated panels of{ size}
$1.5{\,}\rm m\times1.5{\,}\rm m$ around the feed cabin, we
{invert} the potential of the antennae {to
be }able to trace the object. For example, {it
can} detect{ a} fast radio burst (FRB) event with
two peaks (FRB 200428)~\citep{paper:17} and {a}
pulsar in local group M31 with $3.6\times10^5$ s of
observing time by{ dynamically}
deformin{g} {its} elliptical
shape~\citep{paper:18}. Interestingly, we realized
that one of the unique{ aspects} of FAST, the
cross-section{al} area of the whole paraboloid $A_{\rm F}$,
motivated us to reconceive the
{shape} of FAST from geometrical
iteration: we assume $2^{\rm nd}$ order Koch fractals
relying on the iterated function system (IFS) technique. Considering that
the observed structure of the antenna{,} located in the local
universe{,} satisfies the mechanical similarity of{ a} Julia
set~\citep{paper:19}, we build an affine transformation set within
{a} complex space which {is}
similar {to} the process of{ altering} the
sectional area \emph{A}, i.e., the Hutchinson operator used in
fractal scaling \begin{equation}
\mathcal{H}(A)=\bigcup_{k=1}^{m}w_{k}(A)\,. \label{45}
\end{equation}

\begin{figure*}
 \centering
 \includegraphics[width=0.65\textwidth]{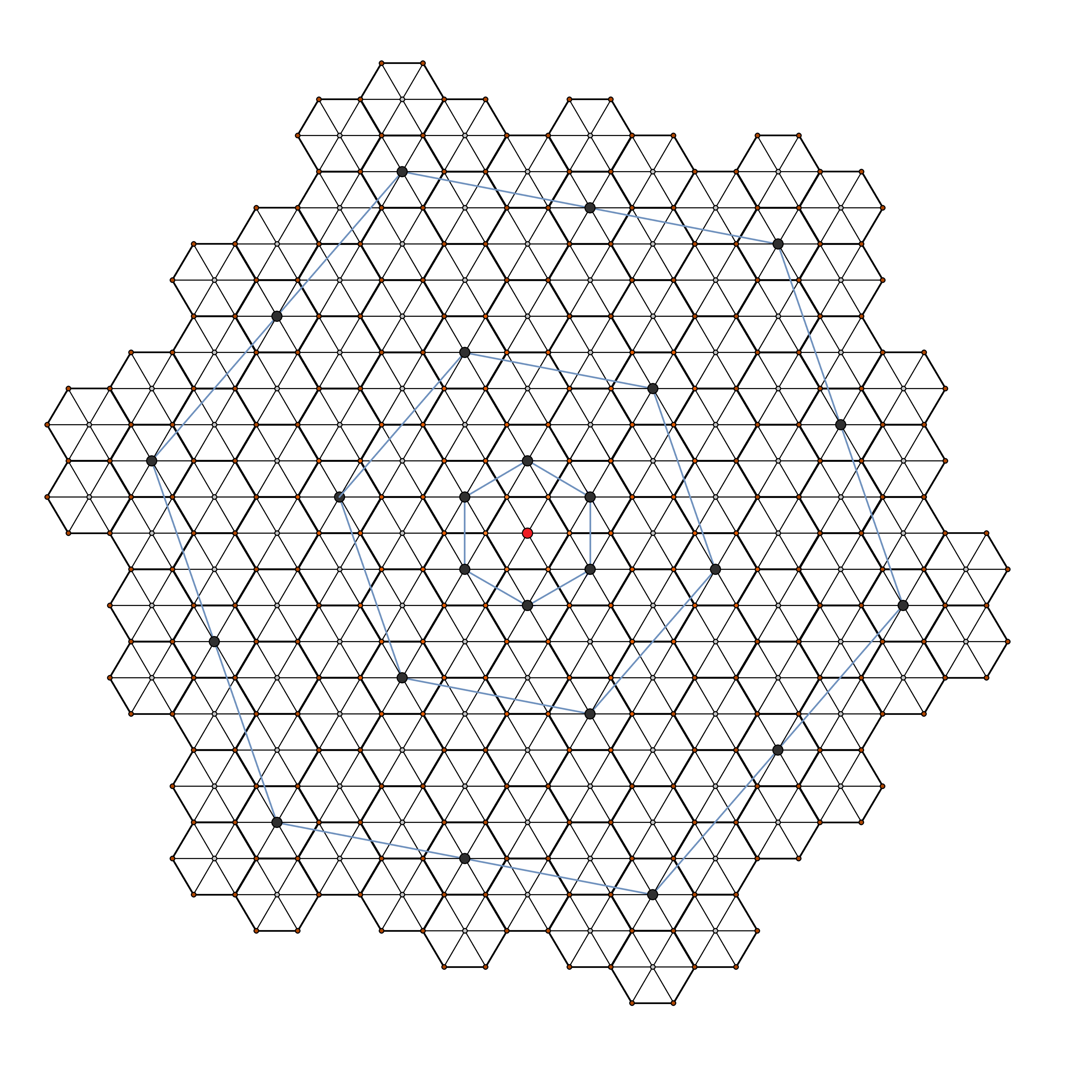}
\caption{\baselineskip 3.8mm The {\it blue-marked edges} in Fig.~\ref{2}
{represent} the partial deformation
f{rom} star tracking. The whole area of FAST is{
composed of} 10{\,}728 {identical}
triang{u}l{ar} shape{d} reflectors. Each actuator
node{ can be} represented in Cartesian coordinates. Considering
the mechanical support, we claim the local spatial coordinates for
every cable element of FAST $\mathscr{Q}_{\rm
FAST}=\rm[\textbf{X,Y,Z}]^T$, where these matrices $\rm\textbf{X}$,
$\rm\textbf{Y}$ and $\rm\textbf{Z}$ are sets of actuator node
positions.}
 \label{2}
 \end{figure*}
We introduce the Hohlfield-Cohen-Ramsey principle (HCR)
\citep{paper:20}, which {describes} the
sufficient conditions for all frequency-independent antennas. The
nature of{ the} transformation produces \begin{equation} (\alpha
A)\bigcap A=\alpha A,\alpha\in\mathbb{R}\,. \label{46}
\end{equation} During the dimensional
continuity {stated} in Equation~\eqref{45}, we
define the coordinate spacing
 \begin{equation}
\mathfrak{Z}=|{\rm z}_i-{\rm z}_j|,\quad i\neq j\,, \label{47}
\end{equation} {in which }each {$z$} denote{s}
the coordinate of the actuator, and the total length of the
broken line described by the proportion increases \begin{equation}
\mathcal{L}=\left(\frac{4}{3}\right)^n\mathfrak{Z}\,. \label{48}
\end{equation}

Referring{ to} research in 2003, K.~J.~Vinoy et al. first
evaluate{d} the relationship between the size of{ an} antenna and
generalized Koch fractal $D$ {with respect to} the ang{u}l{ar}
dependence~\citep{paper:21} \begin{equation} D=\frac{\log
n}{\log(1/\xi)}\,{.} \label{49} \end{equation} {T}he notation
$\xi\in\rm \mathbb{R}$ is the scale factor of the snowflake{,} which
is dependent upon the angle \begin{equation}
\frac{1}{\xi}=\frac{1}{2(1+\cos\theta)}\,. \label{50} \end{equation}
In this case, we substitute $n=4$, and let each edge angle of FAST
{consist of a} unit triang{u}l{ar} board {with}
$\theta=60^{\circ}${, which can be described by}
Equation~\eqref{49}. Thus, each{ affected} actuator {obeys} the
relation \begin{equation} \begin{aligned}
D&=\frac{\log 4}{\log\{2(1+\cos(\pi/3))\}}\\
&=\frac{\log 4}{\log 3}\approx1.26185951\,.
\label{51}
\end{aligned}
\end{equation}

\noindent For {a} pentagon{al} deformation, according to the
practical investigation of FAST, these angles are
$\theta_1=54^{\circ}$ and $\theta_2=72^{\circ}$. Especially{ for}
the junction between the local main reflectors, their dimensions
through the Koch curve calculation are {expressed} {as}
\begin{equation} (D_1,D_2)\approx(1.19974868,0.229966453).
\label{52} \end{equation}

From the point of view of the least disturb{ance} on the
performance of the fractal antenna \begin{equation}
\epsilon(D,D_1)<\epsilon(D,D_2)\,, \label{53} \end{equation} the
surface deformation at the angle of 54 degrees may
{have} the minimum error $\epsilon$. On the other
hand, one of the potentials that can be explored in FAST is the
geometrical curve-fitting approach, and it is
{utilized} to obtain an approximate formula
{f}o{r} {the }design, which belongs to
the area of recursive IFS. From the perspective of power
electronic{s}, we demonstrate one of the original ideas
of the mechanism, for the input impedance through the
fractalization \begin{equation} Z_{f}=Z_{\rm
in}\left[1-(1+0.9\log4)\frac{\log D}{D}\right]^2{.}
\label{54} \end{equation} {S}ubstituting $D_1$ into
Equation~\eqref{54}, we obtain the impedance to express the
fractalized resistance \begin{equation}
Z_{f}\approx0.807\cdot Z_{\rm in}\,. \label{55} \end{equation}

{As} the resistivity of the main-metallic
aluminium is about 2.65 ohm{\,}km{$^{-1}$} per square
millimet{e}r cross-sectional area, we estimate a
rough value of the fractalized resistance is 2.13855
ohm{\,}km{$^{-1}$} per square
millimet{e}r. {Upon comparison},
the{ value} {decreases} by 19.3\%.
Th{is} case shows that the iteration in{ calculating
the} $2^{\rm nd}$ order Koch snowflake may develop to detect{ a}
low-frequency event in SETI backend improvement. We suppose
that the fatigue stress tensor of the local cable network is within
{a} reasonable range of{ the} Miner criterion, and
the adjusted shape of the reflecting surface does not violate the
limitation of proportional-integral-derivative (PID) control
performance of hydraulic actuators. {It is
o}nly necessary to change the axial position of several
groups of actuator nodes of the cable net slightly, so that the
tracking area of the pulsar will generate a shape of up to $2^{\rm
nd}$ order{ for the} aluminium{ panels}
{composing the} Koch fractal antenna. The
corollary is that the {illustration
depicted} in Figure~\ref{2}{ can be referenced to}
develop a strateg{y}{ utilizing}
sensitivity of fractal-area calibration (FAC).

\section{Discussion} Since the main{ }reflector of
FAST is a deformable receiv{er} composed of
2000 actuator units{,} {u}sing the finite Koch
snowflake antenna iteration, we {investigate} the
feasibility of applying it to {a} radio telescope such
as FAST, aiming at optimizing the sensitivity and the observation
capability for very long baseline interferometry (VLBI).
From the ellipticity of EIRP to the influence of sensitivity
equation for {t}echnosignatures, we
{incorporate various} aspects, including the
assumption of {a} second-order fractal antenna for the
estimation of different intersection angles to find {a}
suitable deformation of reflectors. We hope to
appl{y} {a} fractal antenna in
the future and try to implement {such correction to}
FAST, to optimize {observations for}
SETI. Even if {such a
signal} {is like} looking for a needle in a haystack, we
believe it will improve additional opportunities f{or
possible} {exobiological} communication.

\begin{acknowledgements} This work was supported by
the National {Natural }Science Foundation of China (Grant No.
11929301){ and} National Key R\&D Program of China
(2017YFA0402600). This theoretical project was technical{ly}
supported by the FAST work{ing} team, {at} National
Astronomical Observatories, Chinese Academy of Sciences
(NAOC) and the technical lead Ye-Nan Cui from Beijing DEVIN
Technology Co, Ltd. \end{acknowledgements}

\appendix

\section{} Define a Lie group acting on {an}
electromagnetic field \begin{equation}
 L[\boldsymbol{\mathfrak{E,H}}]=L[\boldsymbol{\mathfrak{E}}]\otimes L[\boldsymbol{\mathfrak{H}}]\,.
 \label{55}
\end{equation} According to the work by A.{ }Bayliss \& E.{
}Turkel~\citep{paper:22}, Equation~\eqref{2} has been extended to
spherical coordinates{.} {I}n such a case, the radiation $u$ is
analytic, uniform and convergent if $n>1${.} {T}he series mentioned
in reference~\cite{paper:09} is convergent, {and can be }obtained{
in} a spherical form \begin{equation} \begin{aligned}
R^mu(d,\theta,\varphi)=&\sum_{n=1}^{m}R^{l-n}s_n(d,\theta,\varphi)\\
&+\sum_{n=m+1}^{\infty}R^{l-n}s_n(d,\theta,\varphi)\,,
 \label{57}
 \end{aligned}
\end{equation} where $R=\alpha d$ and $\alpha\in \mathbb{N}$,
multiplying by the order of $1/R$ to the radius. In the case of
spherical harmonics, we express the signal function{
as}~\citep{paper:22} \begin{equation}
 \begin{aligned}
s_n(\theta,\varphi)=&\left[\frac{1}{n!(2j)^n}\right]\cdot\prod_{i=0}^{n}\left\{i(i-1)+\left[\frac{1}{\sin\theta}\frac{\partial}{\partial\theta}\right.\right.\\
&\left.\left.+\left(\sin\theta\frac{\partial}{\partial\theta}\right)+\frac{1}{\sin^2\theta}\frac{\partial^2}{\partial\varphi^2}\right]\right\}s_0(\theta,\varphi).
\label{58}
  \end{aligned}
\end{equation} Once the radial motion is
{asserted}, the ERP on {a} 2D reference frame is the extension
of {electromagnetic} circular polarization.

\section{} For the first stage receiver of{ a} THz
radar, we {have} the {as}sumption: in
{a} large-scale displacement survey, we intuitively
regard the uncertainty principle in a homogeneous universe fits the
narrowband approximation \begin{equation}
\log\left(\frac{c+\textbf{v}}{c-\textbf{v}}\right)
=\sum_{n=0}^{\infty}\frac{2}{2n+1}\cdot\left(\frac{\textbf{v}}{c}\right)^{2n+1}\thickapprox\frac{2\textbf{v}}{c}\,,
 \label{59}
\end{equation} where the Doppler frequency \begin{equation} v_{\rm
D}=-\frac{2\textbf{v}}{\lambda}\,{.}
 \label{60}
\end{equation} {T}he spectr{al} energy can be
written as \begin{equation} \begin{aligned}
\frac{2v}{c}<<&\left\{\left[\frac{\int(v_{0}-v_{\rm cut})^2|s_{1}(v)|^2{\rm d}v}{\int|s_{1}(v)|^2{\rm d}v}\right]\right.\\
&\left.\times\left[\frac{\int(t_{0}-t_{\rm cut})^2|s_{1}(t)|^2{\rm
d}t}{\int|s_{1}(t)|^2{\rm d}t}\right]\right\}^{-1}. \label{61}
\end{aligned} \end{equation} {W}e represent in short
\begin{equation} \left|\frac{v_{\rm D}}{v_{\rm cut}}\right|<<\delta
v
 \label{62}
\end{equation} while the sampling loss of the narrowband
\begin{equation} \delta v\cong\frac{1}{\Delta v\Delta t}
 \label{63}
\end{equation} is neglected. On the other hand, the normalized
ambiguity function of{ a} Doppler constant for narrowband
will be \begin{equation} \frac{\iint_{-\infty}^{+\infty}\left|\int
s_{2}(t)s_{1}^*(t-\tau){\rm e}^{2\pi jvt}{\rm d}t\right|^2{\rm
d}\tau{\rm d}v}{\left|\int s_{1}(t)s_{1}^*(t){\rm d}t\right|^2}=1.
 \label{64}
\end{equation} If the echnosignature is
multi-point noise, we may intuitively see it as the geometrical
cent{e}r of the spectrum \begin{equation} \Delta
v=\frac{\sum_{i}B_{i}\Delta v_{i}{\rm a}_{i}}{\sum_{i}B_{i}{\rm
a}_{i}}\,,
 \label{65}
\end{equation} where the delay frequency{ is} $\Delta
v_{i}${,} corresponding to each R/T site from the target
connection to the area, $B_{i}$ is the component
weight and ${\rm a}_{i}$ is the delay amplitude for each
site. {In a}
non-relativistic{ regime} \begin{equation} \Delta
t=\frac{\sum_{i}B_{i}\Delta t_{i}{\rm a}_{i}}{\sum_{i}B_{i}{\rm
a}_{i}}\,.
 \label{66}
\end{equation} Overall, we define the response H and conveniently
give the orthogonal expression of the ENBW for ideal IFA in the
receiver \begin{equation} B=\frac{1}{|{\rm H}(v_{\rm
max})|^2}\int_{0}^{\infty}|{\rm H}(v)|^2{\rm d}v\,.
 \label{67}
\end{equation}

\section{}
See Table~\ref{C1}.
\begin{table}
\centering
  \caption[]{\centering GLOSSARY}\label{C1}

  \renewcommand\baselinestretch{1.3}
  \fns
    \begin{tabular}{ll}
    \hline\noalign{\smallskip}
  {S}ymbol&{M}eaning\\
  \hline\noalign{\smallskip}
  $\omega$ & Angular frequency{ (}rad sec{$^{-1}$)}\\
 $k_{\rm B}$ & Boltzmann's constant{ (}$1.38\times10^{-23}${\,}J{\,}K{$^{-1}$)}\\
   $A$ & Collecting area of the antenna{ (}$\rm m^2${)}\\
   $v$ & Frequency{ (}Hz{)}\\
    S & Flux density\\
   $G$ & Gain \\
  $\theta,\varphi$ & Intersection angle\\
     $\delta v$ & Narrow bandwidth \\
     $N_{\rm b}$ & Number of baselines \\
    $t$ & Observing time{ (}s{)} \\
    $\eta$ & Observation efficiency \\
   $\Delta v$ & Observing channel bandwidth \\
   $d$ & Radius path or distance{ (}m{)}\\
    $T_{\rm sys}$ & System noise temperature{ }(K)\\
  $c$ &Speed of light{ (}$2.99\times10^8${\,}m{\,}s{$^{-1}$}{)}\\
F & Sensitivity{ (}$\rm m^2{\,}K{^{-1}}${)} \\
   $S_{\rm P}$ & Signal power \\
 $\textbf{v}$ & Velocity{ (}m{\,}s{$^{-1}$)} \\
  $\lambda$ & Wavelength \\
ENBW & Equivalent noise bandwidth\\
ENR & Excess noise ratio\\
 EIRP & Effective isotropic radiated power\\
HCR & Hohlfeld-Cohen-Rumsey principle\\
 IF & Intermediate frequency \\
 IFS & Iterated function system \\
 NF & Noise figure \\
 NPF & Nonlinear phase factor\\
 PRF & Pulse repetition frequencies\\
 RSL & Receive{d} signal {l}evel \\
SEFD & System equivalent flux density\\
SNR & Signal noise ratio \\
    \hline
 \end{tabular}
\end{table}

\label{lastpage}
\end{document}